\documentclass[%
 reprint,
showpacs,preprintnumbers,
 amsmath,amssymb,
 aps,
]{revtex4-1}

\usepackage{graphicx}% Include figure files
\usepackage{dcolumn}% Align table columns on decimal point
\usepackage{bm}% bold math
\usepackage[pagebackref=false,colorlinks,linkcolor=blue,filecolor=green,urlcolor=blue ,citecolor=blue]{hyperref}

\begin{document}

\title{Revival of superconductivity in a one-dimensional dimerized diamond lattice}%  \\
\author{Sanaz Shahbazi and Mir Vahid Hosseini}
 \email[Corresponding author: ]{mv.hosseini@znu.ac.ir}
\affiliation{Department of Physics, Faculty of Science, University of Zanjan, Zanjan 45371-38791, Iran}

\begin{abstract}
We study an s-wave superconductivity in a one-dimensional dimerized diamond lattice in the presence of spin-orbit coupling and Zeeman field. The considered diamond lattice, comprising of three sublattices per unitcell and having flat band, has two dimerization patterns; the intra unitcell hoppings have the same (opposite) dimerization pattern as the corresponding inter unitcell hoppings, namely, neighboring (facing) dimerization. Using the mean-field theory, we calculate the superconducting order parameter self-consistently and examine the stability of the superconducting phase against the spin-orbit coupling, and Zeeman splitting, dimerization, and temperature. We find that the spin-orbit coupling or Zeeman splitting individually has a detrimental effect on the superconductivity, mostly for the facing dimerization. But their mutual effect revives the superconductivity at charge neutrality point for the facing dimerization.
\end{abstract}

%\pacs{ }
\maketitle

%%%%%%%%%%%%%%%%%%%%%%%%%%%%%%%%%%%%%%%%%%%%%%%%%%%%%%%%%%%%%%%%%%%%%%%%%%%
\section {Introduction} \label{s1}
%%%%%%%%%%%%%%%%%%%%%%%%%%%%%%%%%%%%%%%%%%%%%%%%%%%%%%%%%%%%%%%%%%%%%%%%%%%

Superconductivity is an amazing quantum phenomenon in macroscopic scales in which electrons at the Fermi level become unstable against attractive interactions mediated by bosonic fields \cite{TheorySuper}. This instability gives rise to the formation of the so-called Cooper pairs predicted by Bardeen, Cooper and, Schrieffer and known as the BCS theory \cite{BCS}. The search for superconducting states has attracted much interest recently, developing this field to non-BCS superconductivity \cite{HeavyFermion,HighTc,IronSuper} with unconventional pairing symmetries \cite{Unconvensuper,UnconvenPairing}. In the usual Cooper pairing, owing to the large Fermi surface, the lattice structure and, to some extent, the dimensions of host materials have less effects in establishing superconductivity \cite{Tinkham}. However, the formation of exotic forms of superconductivity has been proposed theoretically and realized experimentally in new states of matters \cite{TopoSuper1,TopoSuper2,OddFreq} with unusual lattice structure in low dimensional systems \cite{LowSuper1,LowSuper2}, particularly, in one-dimensional (1D) systems \cite{1DSuper1,1DSuper2}.

Furthermore, superconductivity can be engineered by the spin-orbit interaction \cite{SOSuper} and/or the Zeeman field \cite{ZeemanSuper1,ZeemanSuper2}. Spin-orbit interaction that couples the momentum of an electron to its spin \cite{SO}, has a significant effect on spintronics \cite{spintronics1,spintronics2,spintronics3,spintronics4}. This coupling is a key gradient in the emergence of nontrivial phases \cite{SOTOPO}. Spin-orbit coupling with an external origin is the Rashba spin-orbit interaction \cite{Rashba}, which can be created by applying an electric field perpendicular to the plane of materials through breaking inversion symmetry. Rashba spin-orbit coupling splits spin states into chiral states leading to several physical phenomena including quantum spin-Hall effect, spin transistor, and chiral magnonics \cite{RashbaRev}. Chiral symmetric systems \cite{10} such as Rashba nanowire systems \cite{RashbaWire} and Kitaev chain \cite{Kitaev,TOPOSuper} are needed to study topological superconductor.
Rashba nanowire systems can host Majorana fermions \cite{14,16}. Also, the Zeeman field splits spin states into spin-polarized states causing the pair breaking for the s-wave superconductivity \cite{17} and realizing the Fulde–Ferrell–Larkin–Ovchinnikov (FFLO) state \cite{FFLO}.

On the other hand, the lattice structure of a system along with its distortions and dimensions can be encoded in the quantum states of the band structure \cite{QuantMater}. As such, the physical properties of the system, including superconductivity as well as the dynamics of carriers, governed by the band structure, can be affected by the lattice structure. There are some 2D bipartite lattices with specific geometries, such as Dice, Kagome, and Lieb lattices \cite{flatbandLattices}, having internal symmetries, where the rim sublattices are connected indirectly through hub sublattices. In these lattices, an extra non-dispersive band, i.e., flat band, will be emerged in contrast to the usual dispersive conduction and valence bands. Such flat-band systems can be engineered by implementing dimerization \cite{2Ddimerization}. There are several 1D models, e.g., the 1D diamond lattice, exhibiting flat band in their band structure \cite{1DFlatLattice1,1DFlatLattice2,1DFlatLattice3} that also have been designed experimentally \cite{1DFlatLatticeExp}.

Because of the flat bands, highly correlated phases, e.g., superconductivity, would be established in flat-band systems. Superconductivity in 3D and 2D systems supporting flat \cite{FlantSuper0,FlantSuper1,FlantSuper2,FlantSuper3,FlantSuper30,FlantSuper4,FlantSuper5,KagomeSuper1,KagomeSuper2,KagomeSuper3} or partially flat \cite{intriSupParFlatEnhac1,intriSupParFlatEnhac2,intriSupParFlatEnhac3,intriSupParFlatEnhac4,intriSupParFlatEnhac5,intriSupParFlatEnhac6,intriSupParFlatEnhac7,intriSupParFlatEnhac8} band has been studied extensively, with intrinsic \cite{intriSupFlatQuMet} and extrinsic \cite{ExtriSupFlatDice} origins. The pairings of fermions \cite{PairingFermion} and Cooper pairs \cite{PairingCooper1,PairingCooper2} on a 1D diamond chains embedded in a magnetic field have been studied. Also, the possibility of high-T$_C$ superconductivity has been investigated on a cross-linked ladder \cite{crossLadderSuper}. A considerable binding energy for Cooper pairs has been obtained slightly below 1/3-filling in repulsive interacting fermions on the diamond lattice \cite{SuperDiamond}. Also, nontrivial phases have been revealed in interacting bosons within a Bose-Hubbard model on a cross-linked ladder with $\pi$ flux \cite{crossLaddersuperfluid1,crossLaddersuperfluid2}. It has been shown that superconductivity can be dominated over charge order by adding an attractive component on the 1D Creutz ladder with repulsive interactions between spinless fermions \cite{SuperChargCreutz}. But exploring the superconductivity in flat-band systems engineered by Zeeman field \cite{SuperFlatZeeman}, spin-orbit coupling, and lattice dimerizations \cite{SuperDimerLieb} deserves to be investigated further particularly, in 1D systems.

In this paper, we consider a 1D spin-orbit-coupled diamond lattice with lattice dimerization subjected to the Zeeman field in the presence of an s-wave superconductivity. In the normal state, we find that although the spin-orbit coupling, the Zeeman field, or the dimerization cannot individually affect on the dispersion-less property of the flat band, but their combined effect changes some dispersion-less states at the flat band into nearly dispersive ones. The made dispersion in the flat band depends on the dimerization configuration. In the superconducting state, interestingly, we reveal that although the spin-orbit coupling, the Zeeman field, or the dimerization individually can have detrimental effect on the superconductivity, but their combination revives the superconductivity for a certain dimerization pattern.

The paper is organized as follows. In Sec. \ref{s2}, we present the Hamiltonian of the system and discuss its band structure. We incorporate an attractive interaction for establishing superconductivity and derive gap equation using the mean-field formalism in Sec. \ref{s3}. Section \ref{s4} presents the obtained numerical results. Finally, Sec. \ref{s5} is devoted to summarizing and concluding remarks.

\begin{figure}[h]
    \centering
     \includegraphics [width=8cm]{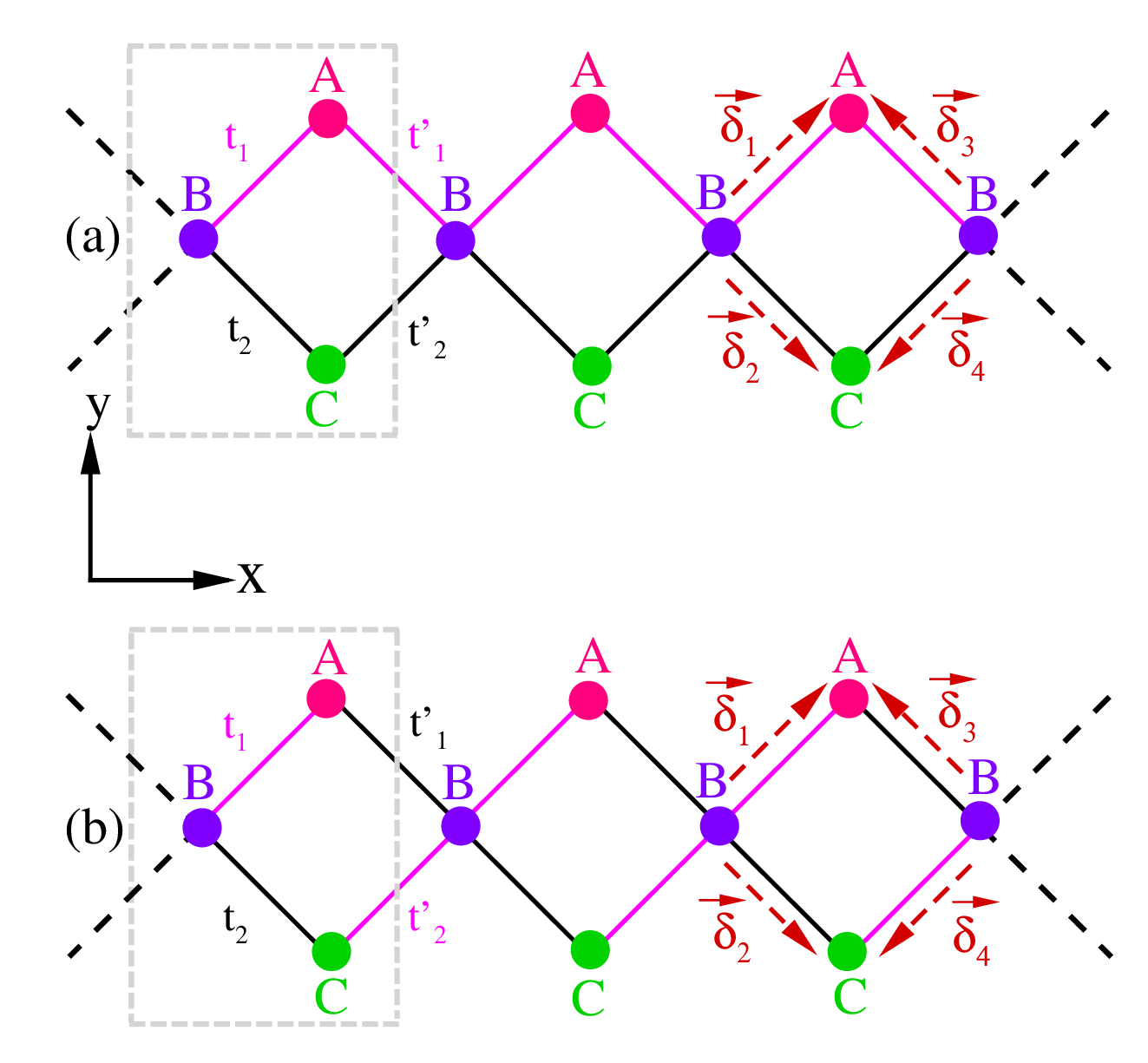}
    \caption{(Color online) Two dimerized configurations of 1D diamond lattice: (a) Neighboring dimerization: The intra and inter unitcell hoppings $B-A$ (or $B-C$) are the same. (b) Facing dimerization: The intra and inter unitcell hoppings $B-A$ (or $B-C$) are the opposite. The dashed box indicates the unitcell.}
        \label{fig1}
\end{figure}
%%%%%%%%%%%%%%%%%%%%%%%%%%%%%%%%%%%%%%%%%%%%%%%%%%%%%%%%%%%%%%%%%%%%%%%%%%%
\section {Model and Theory}\label{s2}
%%%%%%%%%%%%%%%%%%%%%%%%%%%%%%%%%%%%%%%%%%%%%%%%%%%%%%%%%%%%%%%%%%%%%%%%%%%
We consider a 1D diamond lattice along the x axis, as shown in Fig. \ref{fig1}, containing three distinct sublattices (namely, $A$, $B$, and $C$) per unitcell in the presence of the spin-orbit coupling and the Zeeman field. The lattice is also dimerized in two different ways \cite{DimerizedDiamond}: (i) the intra and inter unitcell hoppings $B-A$ (or $B-C$) have the same dimerization, i.e., the neighboring dimerization [see Fig. \ref{fig1}(a)], and (ii) the intra and inter unitcell hoppings $B-A$ (or $B-C$) have the opposite dimerization, i.e., the facing dimerization [see Fig. \ref{fig1}(b)]. The total Hamiltonian for the system including the Hamiltonians of lattice, $H_K$, the spin-orbit coupling, $H_{SO}$, and the Zeeman field, $H_{Z}$, is
\begin{align}\label{NHam}
H_0=H_K+H_{SO}+H_Z,
\end{align}
with
\begin{align}
H_K&=\sum_{i=1}^N \sum_{\sigma} (t_1 c^{\dagger}_{i,1,\sigma} +t_2 c^{\dagger}_{i,3,\sigma})c_{i,2,\sigma}
 \nonumber \\
+&\sum_{i=1}^{N-1} \sum_{\sigma} (t_1^{\prime} c^{\dagger}_{i,1,\sigma} + t^{\prime}_2 c^{\dagger}_{i,3,\sigma}) c_{i+1,2,\sigma} + H.c. \nonumber \\
 +& \sum_{i=1}^N\sum_{m=1}^3\sum_{\sigma} \mu_m c^{\dagger}_{i,m,\sigma}c_{i,m,\sigma},
\end{align}
\begin{align}
H_{SO}=
&-i\lambda \sum_{i=1}^N \sum_{\sigma,\sigma^\prime}[c^{\dagger}_{i,1,\sigma}(\vec{\tau}\times \hat{d}_{1})_{\sigma\sigma^\prime}+c^{\dagger}_{i,3,\sigma}(\vec{\tau}\times \hat{d}_{2})_{\sigma\sigma^\prime}]c_{i,2,\sigma^\prime}\nonumber \\
 &-i\lambda\sum_{i=1}^{N-1} \sum_{\sigma,\sigma^\prime}[c^{\dagger}_{i,1,\sigma}(\vec{\tau}\times \hat{d}_{3})_{\sigma\sigma^\prime}+c^{\dagger}_{i,3,\sigma}(\vec{\tau}\times \hat{d}_{4})_{\sigma\sigma^\prime}]c_{i+1,2,\sigma^\prime}
+H.c.,
\end{align}
\begin{align}
H_{Z}=-h\sum_{i=1}^N\sum_{m=1}^3\sum_{\sigma}  \sigma c^{\dagger}_{i,m,\sigma}c_{i,m,\sigma},
\end{align}
where $c^{(\dagger)}_{i,m,\sigma}$ is the annihilation (creation) operator for an electron on the sublattices $m=1,2,3$ ($A$, $B$, and $C$) at the $i$th unitcell with spin $\sigma=(\uparrow or \downarrow)$. $t^{(\prime)}_{1}$ and $t^{(\prime)}_{2}$ are the intra (inter) unitcell hoppings of upper and lower bonds, respectively. For the neighboring dimerization $t_1=t^{\prime}_1=t(1+\delta t)$ and $t_2=t^\prime_2=t(1-\delta t)$ [see Fig. \ref{fig1}(a)] and for the facing dimerization $t_1=t^{\prime}_2=t (1+\delta t) $ and $ t_2=t^\prime_1=t (1-\delta t)$ [see Fig. \ref{fig1}(b)] with $t$ and $\delta t$ being the strengths of the hopping and the dimerization, respectively. $\mu_m$ is the chemical potential and the symbol H.c. denotes the Hermitian conjugate of the previous operator. $\lambda$ and $h$ are the spin-orbit coupling and the Zeeman field strengths, respectively. $\vec{\tau}$ is the Pauli vector. Also, $d_{j}$'s ($j=1,2,3,4$) are the unit vectors along the intra ($\vec{\delta}_1$ and $\vec{\delta}_2$) and inter ($\vec{\delta}_3$ and $\vec{\delta}_4$) lattice vectors that are given by
\begin{align}
\vec{\delta}_1={(\frac{\sqrt{2}}{2}a,\frac{\sqrt{2}}{2}a)}, \quad \vec{\delta}_2={(\frac{\sqrt{2}}{2}a,-\frac{\sqrt{2}}{2}a)}, \nonumber\\
\vec{\delta}_3={(-\frac{\sqrt{2}}{2}a,\frac{\sqrt{2}}{2}a)}, \quad \vec{\delta}_4={(-\frac{\sqrt{2}}{2}a,-\frac{\sqrt{2}}{2}a)} ,
\end{align}
with $a$ is the distance between two adjacent lattice points. We choose $t$ and $a$ as the energy unit and the length unit, respectively. In the following, to focus on the role of flat bands, we set $\mu_{(1,2,3)}=0$.

Since the 1D system is along the x axis, the Bloch wave vector $\textbf{k}=(k,0)$ is a good quantum number under periodic boundary conditions. Performing Fourier transformation on the basis of $c_{j,m,\sigma}=\frac{1}{\sqrt N} \sum_{k}e^{i\textbf{k}\cdot \textbf{r}_j}c_{k,m,\sigma}$ and $c^{\dagger}_{j,m,\sigma}=\frac{1}{\sqrt N}\sum_{k}e^{-i\textbf{k}\cdot\textbf{r}_j}c^{\dagger}_{k,m,\sigma}$, the Hamiltonian $H_0$, Eq. (\ref{NHam}), can be written as
\begin{align}\label{HkinsoZ}
H_0=\sum_k {\psi}^{\dagger}_kh_0(k) {\psi}_k,
\end{align}
where ${\psi}^{\dagger}_k=(c_{k,1,\uparrow}, c_{k,2,\uparrow}, c_{k,3,\uparrow},c_{k,1,\downarrow}, c_{k,2,\downarrow}, c_{k,3,\downarrow})^{\dagger}$ and
\begin{align}\label{MoHamilNorm}
h_0(k)= \begin{pmatrix}
h_K(k) & h_{SO}(k)\\
h_{SO}(k)^{\dagger} & h_K(k)\\
 \end{pmatrix}+h_Z.
 \end{align}
Here, we have defined the momentum space Hamiltonian of the diamond lattice as
\begin{align}\label{MoHamilkin}
h_K(k)= \begin{pmatrix}
{\mu}_{1} & s(k) & 0\\
s(k)^{\;*} & {\mu}_{2} & g(k)\\
0 & g(k)^{\;*} & {\mu}_{3}
 \end{pmatrix},
 \end{align}
 where
\begin{align*}
s(k)= t_{1} \exp\left(-i\frac{\sqrt{2}}{2}ka\right)+ t^{\prime}_1 \exp\left(i\frac{\sqrt{2}}{2}ka\right),\\
 g(k)= t_{2}\exp\left(i\frac{\sqrt{2}}{2}ka\right)+ t^{\prime}_2 \exp\left(-i\frac{\sqrt{2}}{2}ka\right),
\end{align*}
and the momentum space Hamiltonian of the spin-orbit coupling as
\begin{align}
h_{SO}(k)= \begin{pmatrix}
0 & \lambda_+(k) & 0\\
\lambda_+(k)^{*} & 0 & \lambda_-(k)\\
0 & \lambda_-(k)^{*} & 0
 \end{pmatrix},
 \end{align}
where
 \begin{align}
     \lambda_{\alpha}(k)=i\sqrt{2} a\lambda \left[\cos \left( \frac{\sqrt{2}}{2}ka\right)+\alpha  \sin\left(\frac{\sqrt{2}}{2}ka\right)\right],
 \end{align}
with $\alpha=\pm$. Also, the momentum space Hamiltonian of the Zeeman field takes the form
\begin{align}
h_Z=h Diag(-1,-1,-1,1,1,1),
 \end{align}
where $Diag(x)$ creates a diagonal matrix.

Although Hamiltonian (\ref{MoHamilNorm}) is not  diagonalizable analytically, but one can obtain analytical spectra for specific cases. For $\lambda=0$ and $h=0$, diagonalizing the Hamiltonian (\ref{MoHamilkin}), yields the eigenvalues of the diamond chain as,
\begin{align}\label{eigHamilkin}
    \epsilon(k)={0, \pm \sqrt{\eta+\xi}},
\end{align}\\
with
\begin{align*}
    {\eta}=t^2_1+t^2_2+{(t^\prime_1)}^2+{(t^\prime_2)}^2,\quad
    {\xi}=2\cos(\sqrt{2}ka)(t_1 t^{\prime}_1+t_2 t^{\prime}_2).
\end{align*}
Explicitly, one can see that the diamond lattice has three bands; two dispersive bands and one flat band at zero energy. For the neighboring dimerization, i.e., $t_1=t^{\prime}_1=t(1+\delta t)$ and $ t_2=t^\prime_2=t(1-\delta t)$ [see Fig. \ref{fig1}(a)], the eigenvalues (\ref{eigHamilkin}) reduce as
\begin{align}
    \epsilon(k)= 0, {\pm 2 \cos(\frac{\sqrt{2}}{2}}ka) \sqrt{t^2_1+ t^2_2},
    \label{eigneigh}
\end{align}
while for the facing pattern, i.e., $t_1=t^\prime_2=t(1+\delta t)$ and $ t_2=t^\prime_1=t(1-\delta t)$ [see Fig. \ref{fig1}(b)], we arrive at,
\begin{align}
\epsilon(k)= 0, \pm
 \sqrt{2[t^2_1 + t^2_2+2 t_1 t_2
\cos (\sqrt{2}ka)]}.
\label{eigfac}
\end{align}
For the non-dimerized case, i.e., $\delta t=0$, Eq. (\ref{eigHamilkin}) can be rewritten as,
\begin{align}
    \epsilon(k)=0, \pm 2t\sqrt{2} \cos (\frac{\sqrt{2}ka}{2}).\label{eignon}
\end{align}
Note, for the neighboring dimerization [Eq. \ref{eigneigh}] and non-dimerization [Eq. \ref{eignon}] cases, the spectrum is gapless and the dispersive bands are similar to Dirac band touching at the Brillouin zone boundaries. While the dimerization opens a gap between the two dispersive bands and the flat band in the facing dimerization case [Eq. \ref{eigfac}]. For the neighboring dimerization, the system has chiral symmetry and can reveal topological phase transition depending on the dimerization values. While, in the facing dimerization case, the system has sublattice symmetry with non-topological properties. However, this system with such dimerization pattern can be turned into topological one in the presence of chiral-symmetry breaking adiabatic pumping \cite{DimerizedDiamond}.
\begin{figure*}[t!]
    \centering
     \includegraphics [width=16cm]{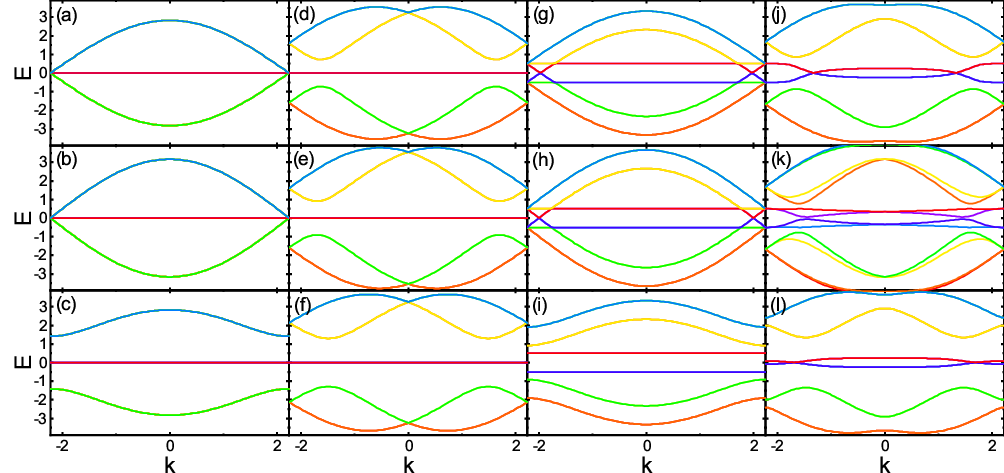}
    \caption{(Color online) The band structure of the system as a function of $k$ for no dimerization (the first row), neighboring dimerization (the second row), and facing dimerization (the third row) patterns. Also, $(\lambda,h)=(0,0)$ for the first column, $(\lambda,h)=(0.8,0)$ for the second column, $(\lambda,h)=(0,0.5)$ for the third column, and $(\lambda,h)=(0.8,0.5)$ for the forth column. Here, $\delta t=0.5$.}
        \label{fig2}
\end{figure*}

The full band structure of the system can be evaluated numerically. In Fig. \ref{fig2}, the band structure versus $k$ is depicted for different cases. The first, the second, and the third rows are for the no dimerization, the neighboring
dimerization, and the facing dimerization patterns, respectively. In the first column, the band structure is calculated in the absence of both $\lambda$ and $h$. The second (third) column is for $\lambda\neq0$ and $h=0$ ($\lambda=0$ and $h\neq0$). The forth column is calculated in the presence of both $\lambda$ and $h$.

From the first column, [see Figs. \ref{fig2}(a), \ref{fig2}(b), and \ref{fig2}(c)], one can see that the no dimerization and the neighboring dimerization have the same gapless band structure including two dispersive bands and one flat band. In these cases, the diamond lattice has Dirac-like bands touching at the 1D Brillouin zone boundaries. While, the facing dimerization opens a gap between the two dispersive bands and the flat band lifting the degeneracy of the Dirac point.

As shown in the second column, the spin-orbit coupling splits the dispersive bands into chiral bands and, at the same time, opens a gap between the dispersive and non-dispersive bands without affecting on the flat band for all the three dimerization patterns [see Figs. \ref{fig2}(d), \ref{fig2}(e), and \ref{fig2}(f)]. The band structures of the three patterns look similar to each other, while the gap of facing dimerization is larger than that of the other two band structures.

As can be seen in the third column, again the band structures of the no dimerization and the neighboring dimerization are the same. In these two configurations, the Zeeman field splits the spin states except at some states close to the Brillouin zone boundaries. In contrast, for the facing dimerization, the Zeeman field lifts the spin degeneracy completely and gaps out the spin states [see Figs. \ref{fig2}(g), \ref{fig2}(h), and \ref{fig2}(i)].

The combined effect of the spin-orbit coupling and the Zeeman field, as depicted in the forth column, results in opening a partial gap in the dispersive band and causing the flat band to acquire dispersion depending on the dimerization patterns [see Figs. \ref{fig2}(j), \ref{fig2}(k), and \ref{fig2}(l)]. Moreover, in the facing dimerization, compared to the other two patterns, the dispersion of the middle bands is smaller and there are more available states near the Fermi energy. It is worthwhile noting that in bipartite lattices, band crossing points and the flatness of the flat band are protected by topological mechanism \cite{FlatTopo}. In the diamond lattice, the zero-energy states result from the absence of direct connection between $A$ and $C$ sublattices. This implies that the corresponding wave function is localized at $A$ and $C$ sublattices with opposite amplitudes and localized at B sublattices with zero amplitude. So, the removed band touching points and the distortion of the flat band can be attributed to the perturbations, i.e., the spin-orbit coupling and the Zeeman field, that do not respect the underlying topology \cite{FlatTopo}.

%%%%%%%%%%%%%%%%%%%%%%%%%%%%%%%%%%%%%%%%%%%%%%%%%%%%%%%%%%%%%%%%%%%%%%%%%%%
\section {Superconductivity}\label{s3}
%%%%%%%%%%%%%%%%%%%%%%%%%%%%%%%%%%%%%%%%%%%%%%%%%%%%%%%%%%%%%%%%%%%%%%%%%%%

Now in this section, we incorporate an s-wave superconductivity to the 1D diamond chain by including the attractive on-site interaction,
\begin{align}\label{int}
H_{int}=-U \sum_{i}\sum_{m=1}^3 [c^{\dagger}_{i,m,{\uparrow}}c_{i,m,\uparrow}c^{\dagger}_{i,m,\downarrow}c_{i,m\downarrow}],
\end{align}
where $U>0$ denotes the on-site attractive pairing interaction. In the present work, we assume the absence of attraction in the spin-triplet channel. Using the mean-field approximation and taking Fourier transform, Eq. (\ref{int}) can be recast into \cite{Tinkham,GrapheneSuper1}
 \begin{align}\label{intMF}
 H_{int}= \sum_{k}\sum_{m=1}^3 [\Delta_kc^{\dagger}_{ k,m,\uparrow}c^{\dagger}_{k,m,\downarrow}+ \Delta^*_kc_{k,m,\downarrow}c_{k,m,\uparrow}],
\end{align}
where
 \begin{align}\label{OrderMF}
 \Delta_k= -\frac{U}{3} \sum_{m=1}^3 \langle c_{k,m,\downarrow} c_{k,m,\uparrow} \rangle, %and $\Delta^*_i= -\frac{U}{3} \sum_{m=1}^3 \langle c^\dagger_{i,m,\uparrow} c_{i,m,\downarrow}\rangle$
 \end{align}
is the mean-field superconducting order parameter. We assume that the correlation functions $\langle c_{k,m,\downarrow} c_{k,m,\uparrow} \rangle$ are the same for all three sublattices $m=1,2,3$ \cite{GrapheneSuper1,GrapheneSuper2,GrapheneSuper3}. Also, in the s-wave pairing $\Delta_k= \Delta^*=\Delta$.

Adding Eq. (\ref{intMF}) to Eq. (\ref{HkinsoZ}), gives the total Hamiltonian $H=H_0+H_{int}$ in the momentum space as,
\begin{align}\label{HT}
H=\sum_k {\Psi}^{\dagger}_k h(k) {\Psi}_k,
\end{align}
with the Nambu spinor
\begin{align}\label{HT}
{\Psi}^{\dagger}_k&=(c_{k,1,\uparrow}, c_{k,2,\uparrow}, c_{k,3,\uparrow},c_{k,1,\downarrow}, c_{k,2,\downarrow}, c_{k,3,\downarrow})^{\dagger}\nonumber\\
\oplus&(c_{k,1,\downarrow}, c_{k,2,\downarrow}, c_{k,3,\downarrow},c_{k,1,\uparrow}, c_{k,2,\uparrow}, c_{k,3,\uparrow})^{\dagger},
\end{align}
and the momentum space total Hamiltonian
\begin{align}\label{MoHamilTotal}
h(k)= \begin{pmatrix}
h_0(k) & \hat{\Delta}\\
\hat{\Delta}^{*} & -h_0(k)^T\\
 \end{pmatrix},
 \end{align}
 where
 \begin{align}
\hat{\Delta}=\frac{\Delta}{2} Diag(1, 1, 1,-1,-1,-1).
 \end{align}
Invoking the Bogoliubov-Valatin transformation \cite{Bogoliubov-Valatin,BlackS,JacobL},
\begin{align}\label{Bogoliubov-Valatin}
c_{k,m,\sigma}=\sum_{\nu}    (u_{k,m,\sigma}^{\nu}\gamma_{\nu}+ v_{k,m,\sigma}^{\nu*}\gamma_{\nu}^{\dagger}),
\end{align}
Hamiltonian (\ref{MoHamilTotal}) can be diagonalized by solving
\begin{align}
h_T(k) \psi^{\nu}_k= E^{\nu}(k)\psi^{\nu}_k,
\label{BdGeq}
\end{align}
where $E^{\nu}(k)$ are the eigenvalues and
\begin{align}
\psi^{\nu}_k&=(u^{\nu}_{k,1,\uparrow}, u^{\nu}_{k,2,\uparrow}, u^{\nu}_{k,3,\uparrow},u^{\nu}_{k,1,\downarrow}, u^{\nu}_{k,2,\downarrow}, u^{\nu}_{k,3,\downarrow},\nonumber\\
&v^{\nu}_{k,1,\downarrow}, v^{\nu}_{k,2,\downarrow}, v^{\nu}_{k,3,\downarrow},v^{\nu}_{k,1,\uparrow}, v^{\nu}_{k,2,\uparrow}, v^{\nu}_{k,3,\uparrow})^T,
\end{align}
are the eigenvectors of the system. Here, $u_{k,m,\sigma}^{\nu}$ and $v_{k,m}^{\nu}$ are the electron and hole states, respectively. Also, $\gamma_{\nu \sigma}^{\dagger}(\gamma_{\nu \sigma})$ is the quasi-particle  creation (annihilation) operator in the $\nu$ state with spin $\sigma$. Plugging Eq. (\ref{Bogoliubov-Valatin}) into Eq. (\ref{OrderMF}), one obtains the superconducting gap equation as
\begin{equation}
\Delta=\frac{U}{3}\sum_{k,\nu}\sum_{m=1}^{3}u_{k,m,\downarrow}^{\nu}v_{k,m,\uparrow}^{\nu*} \tanh\left[\frac{E^\nu(k)}{2k_B T}\right],
\label{gapEign}
\end{equation}
where $T$ is the temperature and $k_B$ is the Boltzmann constant. With an initial guess for the order parameter $\Delta$, one can solve the eigenvalue problem (\ref{BdGeq}). Having obtained the eigenvalues and the eigenvectors of the system and setting them into the gap equation (\ref{gapEign}), one can determine a new value for $\Delta$. This process can be done iteratively obtaining the order parameter self-consistently.

To examine the stability of superconducting phase, the calculated $\Delta$ should minimize the thermodynamic potential \cite{Thermo1},
\begin{align}
\Omega_S=-k_BT\sum_{k,\nu}\sum_{\alpha=\pm}\ln\left(1+\exp\left[{\frac{\alpha E^\nu(k)}{k_BT}}\right]\right)+\frac{3\Delta^2}{U},\label{LdFr}
\end{align}
with the global minima. Also, the DOS at zero temperature can be calculated by the following equation,
\begin{align}
DOS(E)=\sum_{k}\sum_{\nu}\delta[E-E^\nu(k)]. \label{DOS}
\end{align}
Note that in Eqs. (\ref{gapEign}) and (\ref{LdFr}) all the positive eigenvalues are summed over \cite{Thermo2}. If we set $\Delta=0$ in Eq. (\ref{LdFr}), the thermodynamic potential of the normal state $\Omega_N$ can be calculated. In order to obtain analytical expressions for some limiting cases, in the following, we replace $\sum_{k}\rightarrow \frac{a}{\sqrt{2}\pi} \int dk$.

In the absence of the dimerization, the spin-orbit coupling, and the Zeeman field, the gap equation (\ref{gapEign}) reads as
\begin{equation}
\Delta=\frac{Ua}{3\sqrt{2}\pi}\int\!\! dk %\sum_{k}%\sum_{\nu=1}^{6}\frac{\partial E^\nu(k)}{\partial \Delta}
\left(\tanh\left[\frac{\Delta}{2k_B T}\right]+\frac{2\Delta}{E(k)}\tanh\left[\frac{E(k)}{2k_B T}\right]\right),
\label{gapEign00}
\end{equation}
where
\begin{equation}
E(k)=\sqrt{\Delta^2+\epsilon_k^2},
\label{Eign00}
\end{equation}
with $\epsilon_k$ being the dispersive band of normal diamond lattice.

The critical temperature $T_c$ can be calculated analytically by setting $\Delta\rightarrow 0$ and  $T\rightarrow T_c$ in Eq. (\ref{gapEign00}). In the low energy limit, that is satisfied at zero doping, and $T_c\rightarrow 0$, the integral of Eq. (\ref{gapEign00}) can be performed easily, yielding,
\begin{equation}
2k_BT_c=t [\mathcal{W}(c^{-1}e^{\frac{3}{2bU}})]^{-1},
\label{Tcc}
\end{equation}
where $\mathcal{W}$(x) is the Lambert $\mathcal{W}$-function, $b=\frac{1}{\sqrt{2} \pi t}$, and $c=\frac{8\gamma}{\pi}$ with $\gamma$ being the Euler's constant. For $U\ll 1$ the above equation can be approximated as
\begin{equation}
k_BT_c\approx\frac{ b Ut}{3-2 b U \ln\ln(c^{-1}e^{\frac{3}{2 b U}})^c}.
\label{TcApprox}
\end{equation}
One can see that the critical temperature $T_c$ is proportional to $U$.

On the other hand, in the absence of both the dimerization and the spin-orbit coupling, the gap equation (\ref{gapEign}) at $T=0$ can be simplified as \cite{ZeroTemGap},
\begin{align}
\frac{3\Delta}{U}=\sum_{k,\nu}\frac{\partial E^{\nu}_{\uparrow}(k)}{\partial \Delta} \Theta(E^{\nu}_{\uparrow}(k)),\label{ZeroGap}
\end{align}
where $E^{1}_{\sigma}=\Delta-\sigma h$, $E^{2,3}_{\sigma}=E(k)-\sigma h$, and $\Theta(x)$ is the Heaviside Theta function. Changing the summation into the integral and performing the integral in the low energy limit yield,
\begin{align}
\frac{3}{2bU}&=\frac{t}{\Delta} \Theta(\Delta-h)+\ln \frac{2t+\sqrt{(2t)^2+\Delta^2}}{\Delta}\nonumber\\
&-\Theta(h-\Delta) \ln \frac{h+\sqrt{h^2-\Delta^2}}{\Delta}.
\label{ZeroGapSolv}
\end{align}
The above equation dictates that there exist two solutions for the gap, namely, the BCS solution ($\Delta_{00}$) if $h<\Delta$ and the Sarma solution ($\Delta_{0h}$) if $h>\Delta$. In either case, one straightforwardly obtains,
\begin{align}
\Delta_{00}&=t [\mathcal{W}(\frac{e^{\frac{3}{2bU}}}{4})]^{-1},\label{ZeroGapBCS}\\
\Delta_{0h}&=\sqrt{\Delta_{00}e^{-\frac{t}{\Delta_{00}}}(2h-\Delta_{00}e^{-\frac{t}{\Delta_{00}}})},
\label{ZeroGapSARMA}
\end{align}
Note that Eq. (\ref{ZeroGapBCS}) for $U\ll 1$ can be approximated as $\Delta_{00}\approx 2tbU/3$, implying that the superconducting gap is proportional to $U$, due to the flat band \cite{FlantSuper0}.

In order to inspect which of the above-mentioned solutions is stable, we evaluate $\Omega_S-\Omega_N$ using Eq. (\ref{LdFr}) at $T\rightarrow 0$. After performing the integration in the low energy limit, one gets
\begin{align}
\Omega_S-\Omega_N&=\frac{b}{3} [2h^2-\Delta_{00}^2+2t(h-\Delta_{00})]\Theta(\Delta_{00}-h)\nonumber\\
&+\frac{b}{3} [2h-\Delta_{00}e^{-\frac{t}{\Delta_{00}}}]^2\Theta(h-\Delta_{00}).
\label{ZeroGapSolv}
\end{align}
The first term, which holds for the Sarma solution, is always a positive quantity. This indicates that the thermodynamic potential of the Sarma superconductivity is larger than that for the normal state. Thus, the Sarma superconductivity is not stable. However, the second term, related to the BCS superconductivity, can be either positive or negative depending on the critical field,
\begin{align}
h_c&=\sqrt{(\frac{t}{2})^2+\frac{\Delta_{00}}{2}(\Delta_{00}+2t)}-\frac{t}{2},
\label{criticalField}
\end{align}
below which the BCS solution is the stable one. The obtained critical field $h_c$ is in contrast to the usual Clogston-Chandrasekhar limit \cite{ClogChandra1,ClogChandra2}. Remarkably, the terms containing $t$ in Eq. (\ref{criticalField}) stem from the existence of the flat band. So, if $t\rightarrow 0$, the Clogston-Chandrasekhar critical field, i.e., $h_c=\Delta_{00}/\sqrt{2}$, can be recovered.

%%%%%%%%%%%%%%%%%%%%%%%%%%%%%%%%%%%%%%%%%%%%%%%%%%%%%%%%%%%%%%%%%%%%%%%%%%%
\section {Numerical results and discussions} \label{s4}
%%%%%%%%%%%%%%%%%%%%%%%%%%%%%%%%%%%%%%%%%%%%%%%%%%%%%%%%%%%%%%%%%%%%%%%%%%%
\begin{figure}[t]
    \centering
     \includegraphics [width=8.5cm]{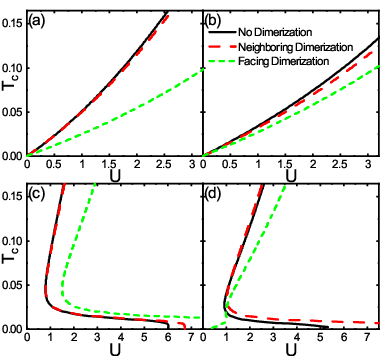}
    \caption{(Color online) Dependence of $T_c$ on $U$ for the no dimerization, neighboring dimerization, and facing dimerization patterns with (a) $(\lambda,h)=(0,0)$, (b) $(\lambda,h)=(0.7,0)$, (c) $(\lambda,h)=(0,0.07)$, and (d) $(\lambda,h)=(0.7,0.07)$. Here, $\delta t=0.5$.}
        \label{Tc}
\end{figure}
\begin{figure}[bh!]
    \centering
     \includegraphics [width=7.5cm]{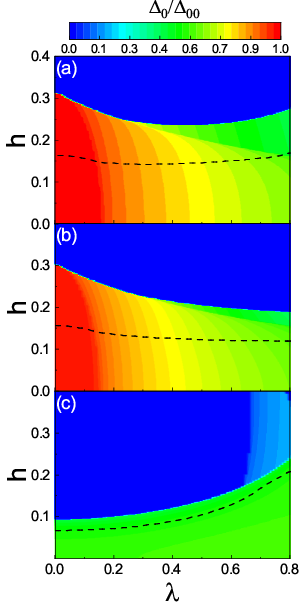}
    \caption{(Color online) Zero temperature phase diagram as functions of $\lambda$ and $h$ for (a) the no dimerization, (b)  neighboring dimerization, and (c)  facing dimerization patterns. The dashed line represents the first order phase transition boundary. $\Delta_0$ is normalized by the superconducting gap $\Delta_{00}$. Here, $\delta t=0.5$ and U=2.5.}
        \label{phaseT0}
\end{figure}

The dependence of critical temperature $T_c$ on the coupling strength $U$ is depicted in Fig. \ref{Tc} for the no dimerization, the neighboring dimerization, and the facing dimerization patterns. As shown in Fig. \ref{Tc}(a), without the spin-orbit coupling and the Zeeman field, interestingly, for small values of $U$ there is a finite value for $T_c$ such that $T_c$ is proportional to $U$. As already discussed, this is because of the existence of flat band at the Fermi level implying the onset of the Cooper pairing even for an infinitesimally small value of $U$ without dispersive bands as well as a finite Fermi surface. Also, remarkably, as $U$ increases, the critical temperature $T_c$ of the facing dimerization remains smaller than those of the neighboring dimerization and the no dimerization patterns. Therefore, Cooper pairing would be weakened due to the facing dimerization. Moreover, the critical temperatures for the neighboring dimerization and non-dimerized case are close together. As a result, the neighboring dimerization and no dimerization are the structures facilitating the Cooper pairing. On the other hand, in the presence of the spin-orbit coupling, as can be seen from Fig. \ref{Tc}(b), the critical temperatures of neighboring dimerization and no dimerization decrease and get closer to that of the facing dimerization compared to Fig. \ref{Tc}(a). As such, for small values of $U$, the $T_c$'s of the three patterns are still proportional to $U$ but their values are almost the same regardless of the dimerization pattern. As a result, the spin-orbit coupling spoils the effect of dimerization. However, for large values of $U$, there is a small deviation between the $T_c$'s of the three patterns. In Fig. \ref{Tc}(c), the critical temperatures are displayed for a finite value of the Zeeman field. Interestingly, one finds that there is a critical value for $U$ below which there is no solution for $T_c$. This means that the formation of Cooper pairs is forbidden. The critical value of $U$ for the facing dimerization is larger than those of the other two patterns. More interestingly, as shown in Fig. \ref{Tc}(d), in the presence of both $\lambda$ and $h$, the quantum criticality of facing dimerization is removed and superconductivity can be established even for small values of $U$. Although, the neighboring dimerization and no dimerization cases were the two favorable structures in Cooper pairing with $\lambda=0$ and $h=0$, but, in this case, they cannot host superconductivity at small values of $U$. In both Figs. \ref{Tc}(c) and \ref{Tc}(d), in some ranges of $U$, there are two critical temperatures due to applying the Zeeman field. The presence of the Zeeman field lifts the spin degeneracy and shifts the spin-subbands. Since Cooper pairs in the s-wave superconductivity are made of two coupled electrons with opposite spins, each of the two electrons lies on different Fermi levels of the spin-splitted subbands. Subsequently, this provides a different Fermi sea for each spin species resulting in the two solutions for $T_c$.

The zero temperature superconducting gap $\Delta_0$ as functions of $\lambda$ and $h$ is plotted in Fig. \ref{phaseT0} with $\delta t=0.5$. $\Delta_0$ is normalized by the zero temperature superconducting gap $\Delta_{00}$ that is calculated in the absence of the spin-orbit coupling, the Zeeman field, and the dimerization. The dashed line indicates the first order phase transition boundary between the normal (upper region) and the superconducting (lower region) phases. In the no dimerization [Fig. \ref{phaseT0}(a)] and the neighboring dimerization [Fig. \ref{phaseT0}(b)] cases, the order parameter $\Delta_0$ is large for small values of both $h$ and $\lambda$. As $\lambda$ increases, the stable $\Delta_0$ decreases almost independent of $h$. The overall values of $\Delta_0$ in the neighboring dimerization [Fig. \ref{phaseT0}(b)] are slightly smaller than those for the no dimerization [Fig. \ref{phaseT0}(a)]. In both figures, the phase transition line is almost a horizontal line with small variations. In contrast, for the facing dimerization [Fig. \ref{phaseT0}(c)], although $\Delta_0$ has smaller values compared to the two previous cases, but the considerable $\Delta_0$ is shifted towards the large $\lambda$. Also, the phase transition line is non-uniform so that the stable superconductivity can sustain even large amounts of fields.

\begin{figure}[t]
    \centering
     \includegraphics [width=8.5cm]{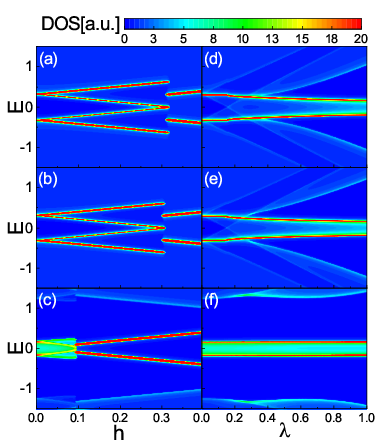}
    \caption{(Color online) Left column: The zero temperature DOS of the system as functions of $E$ and $h$ with $\lambda=0$. Right column: DOS of the system as functions of $E$ and $\lambda$ with $h=0$. The first, the second, and the third rows are for the no dimerization, the neighboring dimerization, and the facing dimerization patterns, respectively. Here, $\delta t=0.5$ and U=2.5.}
        \label{DOS}
\end{figure}

In Figs. \ref{DOS}(a)-\ref{DOS}(c), the DOS of the system versus $E$ and $h$ is depicted with $\lambda=0$, respectively, for the no dimerization, the neighboring dimerization, and the facing dimerization, using the obtained self-consistent solution of the gap equation. For the no dimerization and the neighboring dimerization cases, at $h=0$, there is a superconducting gap around the Fermi level splitting the high density flat band. As $h$ increases, each flat band splits into two diverging bands such that the superconducting bandgap becomes narrower. At a certain value of $h$, since the superconducting gap collapses suddenly, the four high density bands abruptly merge into two Zeeman-splitted bands [Figs. \ref{DOS}(a) and \ref{DOS}(b)]. In contrast, for the facing dimerization case, as shown in Fig. \ref{DOS}(c), only a weak superconducting gap can split the flat band. For small values of $h$, the superconducting gap closes and then two Zeeman-splitted bands reveal with increasing $h$.

The Rashba spin-orbit dependence of the DOS is shown in Figs. \ref{DOS}(d)-\ref{DOS}(f), respectively, for the no dimerization, the neighboring dimerization, and the facing dimerization cases. In the case of the no dimerization and the neighboring dimerization [see Figs. \ref{DOS}(d) and \ref{DOS}(e)], one can see that at small values of $\lambda$, similar to Figs. \ref{DOS}(a)-\ref{DOS}(b), a considerable superconducting gap splits the flat band into two parts. With the increase of the Rashba spin-orbit coupling, the gap between the two high density bands decreases and at the same time the two bands become widen so that a finite DOS can be accessed within the two bands. But, for the facing dimerization case [Figs. \ref{DOS}(f)], the energies of the splitted bands are almost independent of the spin-orbit coupling. Also, there exists a finite value of the DOS between the two high density bands. This implies that a weak superconducting gap is established in this case. Note that, as can be seen from Fig. \ref{DOS}, the DOS is vanishingly small away from the charge neutrality point (flat band). This causes the superconductivity to be declined for all types of the dimerization patterns significantly even in the presence of the spin-orbit coupling.

\begin{figure}[t]
    \centering
     \includegraphics [width=8.5cm]{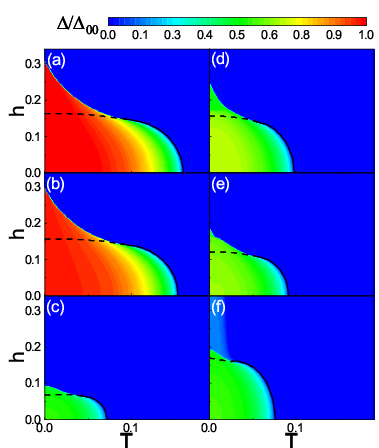}
    \caption{(Color online) Superconducting phase diagram as functions of $T$ and $h$ for $\lambda=0$ (left column) and $\lambda=0.7$ (right column). The first, the second, and the third rows are for the no dimerization, the neighboring dimerization, and the facing dimerization patterns, respectively. The dashed and solid lines represent, respectively, the first and the second order phase transition boundary. $\Delta$ is normalized by the superconducting gap $\Delta_{00}$. Here, $\delta t=0.5$ and U=2.5.}
        \label{phasehT}
\end{figure}

Since the Rashba spin-orbit coupling has smooth effects on the superconductivity at zero Fermi energy, we have investigated the phase diagram in the ($h$,$T$)-plane for zero (left column) and finite (right column) values of the Rashba spin-orbit coupling, shown in Fig. \ref{phasehT}. In the absence of the Rashba spin-orbit coupling, for the no dimerization and the neighboring dimerization [Figs. \ref{phasehT}(a) and Fig. \ref{phasehT}(b)], a considerable $\Delta$ can be obtained over a broad range of the parameters $h$ and $T$. However, the facing dimerization decreases not only the magnitude but also the range of $\Delta$ [Figs. \ref{phasehT}(c)]. In the presence of Rashba spin-orbit coupling, furthermore, both the magnitude and the range of $\Delta$ are decreased in the no dimerization and the neighboring dimerization cases [Figs. \ref{phasehT}(d) and Fig. \ref{phasehT}(e)] implying that the Rashba spin-orbit coupling weakens the superconductivity. Interestingly, as shown in Fig. \ref{phasehT}(f), unlike the two previous configurations, the Rashba coupling along with the facing dimerization promotes the superconductivity, particularly, along the $h$ axis. This is in sharp contrast to the usual cases where the Zeeman splitting has detrimental effects on the superconductivity. Such promotion can be interpreted as follows. As already discussed above, the presence of both Zeeman field and Rashba coupling splits the flat band and, at the same time, makes the band more dispersive as its bandwidth grows. Subsequently, most of the states shift towards higher energies. This decreases available states with large momentum near the Fermi level. As will be shown below, adding the facing dimerization stabilizes the states [see Fig. \ref{fig8}(d)] so that the curvature and the energy states of the middle bands decrease providing low-energy nearly flat band. So, the re-existence of more available states with nearly flat character around the Fermi energy [see Fig. \ref{fig2}(l)] revives superconductivity.

Furthermore, in Fig. \ref{phasehT}, the black dashed and solid lines indicate, respectively, the first and the second order phase transition boundaries between the superconducting and the normal states. The areas below these lines represent a stable superconducting phase where the superconducting thermodynamic potential is less than the thermodynamic potential of the normal states. However, in the normal phase, that is above the dashed line, the superconducting gap can even take non-zero values. This originates from the fact that the superconducting gap is a non-linear equation providing multi solutions such that the stable one resides in the global minimum of the thermodynamic potential. Also, the first-order critical temperature is large compared to the conventional case. This is due to the presence of flat band. Large DOS, provided by the flat band, pairs electrons strongly with a relativity robust superconducting gap. Moreover, quasi-particle excited states are not available just above the superconducting gap [see Fig. \ref{DOS}] and cannot be reached by thermal excitation. So, the Cooper pairs can sustain a relatively large first-order critical temperature.

\begin{figure}[t!]
    \centering
     \includegraphics [width=6cm]{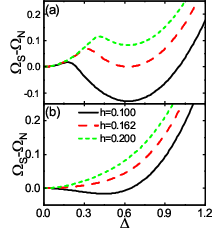}
    \caption{(Color online) Thermodynamic potential difference $\Omega_S-\Omega_N$ as a function of $\Delta$ for various values of $h$ at (a) low temperature $T=0.01$ and (b) high temperature $T=0.11$. Here, $\delta t=0$, $\lambda=0$, and U=2.5.}
        \label{fig7}
\end{figure}
In order to see, how the global minimum of the thermodynamic potential changes either abruptly or smoothly establishing either the first or the second order phase transition when the Zeeman field increases, we have plotted $\Omega_S-\Omega_N$ as a function of $\Delta$ both near zero temperature [Fig. \ref{fig7}(a)] and near the critical temperature [Fig. \ref{fig7}(b)]. As can be seen from Fig. \ref{fig7}(a), $\Omega_S-\Omega_N$ has two minima; a local minimum and a global minimum. For small $h$, the local minimum is located at $\Delta=0$ and the global minimum is at a finite value of the $\Delta$. As $h$ increases, at the critical field $h_c$, the two minima have the same depth. With the further increase of $h$, the thermodynamic potential difference has a lowest value at $\Delta=0$. As a result, the first order phase transition takes place. In contrast, as shown in Fig. \ref{fig7}(b), the thermodynamic potential difference has only one minimum point such that as $h$ increases, this point moves towards $\Delta=0$ gradually. Consequently, the second order phase transition occurs. Note that although Fig. \ref{fig7} is depicted for non-dimerized pattern without the spin-orbit coupling, but the overall behavior is the same for the other patterns even with the spin-orbit coupling (not shown).

\begin{figure}[bh!]
    \centering
     \includegraphics [width=8.5cm]{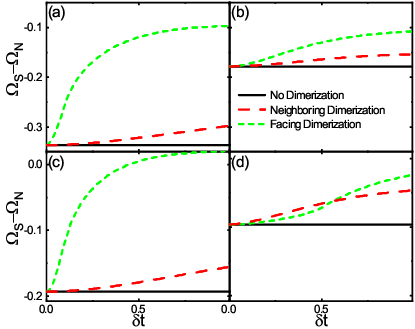}%{fig3.eps}
    \caption{(Color online) Dependence of $\Omega_S-\Omega_N$ on $\delta t$ for the no dimerization, neighboring dimerization, and facing dimerization patterns with (a) $(\lambda,h)=(0,0)$, (b) $(\lambda, h)=(0.7, 0)$, (c) $(\lambda, h)=(0, 0.07)$, and (d) $(\lambda, h)=(0.7, 0.07)$. Here, $T=0$ and U=2.5.}
        \label{fig8}
\end{figure}
In Fig. \ref{fig8}, the thermodynamic potential difference is evaluated by the self-consistent solution of $\Delta$ and depicted versus $\delta t$ for the three structural patterns and various values of ($\lambda$, $h$) at $T=0$. The non-dimerized case has the lowest energy and, obviously, is independent of $\delta t$. But the energies of the neighboring dimerization, and the facing dimerization cases increase with $\delta t$.  Also, the facing dimerization configuration has the highest energies in the absence of both $\lambda$ and $h$ [see Fig. \ref{fig8}(a)] or in the presence of either $\lambda$ [see Fig. \ref{fig8}(b)] or $h$ [see Fig. \ref{fig8}(c)]. However, in the presence of both $\lambda$ and $h$, interestingly, as shown in Fig. \ref{fig8}(d), the facing dimerization has lower energies than those for the neighboring dimerization below a certain value of $\delta t$. So, as already discussed, the superconductivity can be revived in the facing dimerization due to its stabilization via both the spin-orbit and Zeeman field.

Finally, let us comment on the doped case, $\mu_{1,2,3}\neq0$. In this case, the Fermi level resides away from the flat band. Subsequently, a finite Fermi surface establishes with relatively low DOS and the contribution of the flat band to the superconductivity decreases. Therefore, similar to the usual cases, the spin-orbit coupling and the Zeeman field diminish the superconductivity such that, in the facing dimerization, the superconductivity cannot be revived anymore and the results get reduced to the trivial cases.

%%%%%%%%%%%%%%%%%%%%ng %%%%%%%%%%%%%%%%%%%%%%%%%%%%%%%%%%%%%%%%%%%%%%%%%%%%%%%
\section {Summary} \label{s5}
%%%%%%%%%%%%%%%%%%%%%%%%%%%%%%%%%%%%%%%%%%%%%%%%%%%%%%%%%%%%%%%%%%%%%%%%%%%

We considered 1D diamond lattice subjected to the spin-orbit coupling and the Zeeman field posing three structural configurations: the no dimerization, the neighbiring dimerization, and the facing dimerization. We studied normal band structures of the system as well as the dependence of the superconductivity on the lattice structure, temperature, spin-orbit, and Zeeman field. In the normal state, although, individually, either the spin-orbit coupling or the Zeeman field cannot affect the flat band but their combination makes the flat band dispersive. Depending on the type of the lattice configuration, the flat band distortion is different such that for the facing dimerization the flat band remains nearly flat with more available states near the Fermi level. Correspondingly, in the superconducting states, the spin-orbit or the Zeeman field individually has detrimental effects on the superconductivity for each type of the lattice dimerization patterns. But the mutual effect of both the spin-orbit and the Zeeman field would revive the superconductivity in the facing dimerization case. Based on current experimental status, the experimental realization of the system is possible using cold atoms in optical lattices \cite{1DFlatLattice2}, solid-state \cite{1DFlatLatticeExp}, and photonic \cite{ExpPhotoFlat} systems.

\end{document}